\documentclass[11pt,letterpaper]{emulateapj}

\usepackage{graphicx}
\usepackage{dcolumn}
\usepackage{bm}
\usepackage{array}

\usepackage{color}

\newcommand{\be}{\begin{equation}}
\newcommand{\ee}{\end{equation}}
\newcommand{\bea}{\begin{eqnarray}}
\newcommand{\eea}{\end{eqnarray}}
\usepackage{amsmath}
\usepackage{amssymb}

\begin{document}

\title{A method to extract the redshift distortions $\beta$ parameter in configuration space from minimal cosmological assumptions}

\author {Domenico~Tocchini-Valentini}
\email{dtv@pha.jhu.edu, domenico.tocchini@gmail.com}


\author {Michael~Barnard}

\author {Charles~L.~Bennett}

\author {Alexander~S.~Szalay}

\affil {Department of Physics and Astronomy, The Johns Hopkins University, 3400 North Charles Street, Baltimore, MD 21218-2686, USA}

\begin {abstract}
We present a method to extract the redshift-space distortions $\beta$ parameter in configuration space with a minimal set of cosmological assumptions. We show that a novel combination of the observed monopole and quadrupole correlation functions can remove efficiently the impact of 
mild non linearities and redshift errors. The method offers a series of convenient properties: it does not depend on the theoretical linear correlation function, the mean galaxy density is irrelevant, only convolutions are used, there is no explicit dependence on linear bias. Analyses based on dark matter $N$-body simulations and Fisher matrix demonstrate that errors of a few percent on $\beta$ are possible with a full sky, $1\,(h^{-1}\,\mathrm{Gpc})^3$ survey centered at a redshift of unity and with negligible shot noise. We also find a baryonic feature in the normalized quadrupole in configuration space that should complicate the extraction of the growth parameter from the linear theory asymptote, but that does not have a major impact with our method.
\end{abstract}


\maketitle

\section {Introduction} \label {sec:intro}

The gravitational instability picture involves the growth of early time primordial fluctuations into the actual large-scale structure observed through the galaxy distribution. This growth depends in principle on the underlying theory of gravity and the cosmic expansion history. It is therefore important to be able to measure the growth history to obtain useful cosmological informations.

One way of determining the growth of structure is through the apparent anisotropy of the galaxy distribution in redshift space, caused by the line-of-sight (LOS) component of the galaxies peculiar velocities. On large scales, under the linear perturbations regime, the two-point correlation function presents a squashing along the LOS and correspondingly the power spectrum appears enhanced for wavevectors directed along the LOS \citep{1987MNRAS.227....1K}. The anisotropies are governed by the parameter $\beta$ that depends on the growth function and galaxy bias.

Redshift-space distortions (RSDs) have been the subject of many analyses, as reviewed in \cite{1998ASSL..231..185H}. Examples of recent studies involving the latest large surveys are the following. The Two-Degree Field Galaxy Redshift Survey enabled the RSD measurements on the correlation function \citep{2001Natur.410..169P,2003MNRAS.346...78H} and power spectrum \citep{2004MNRAS.353.1201P}. The Sloan Digital Sky Survey permitted other RSD measurements on the correlation function \citep{2005ApJ...621...22Z,2008ApJ...676..889O,2009MNRAS.393.1183C,2009MNRAS.396.1119C} and power spectrum \citep{2004ApJ...606..702T,2006PhRvD..74l3507T}. The VIMOS-VLT Deep Survey and the 2SLAQ Survey were used in \cite{2008Natur.451..541G} for RSD determinations from the correlation function. After this work was submitted RSD studies on the WiggleZ and BOSS catalogs also appeared \citep{2011MNRAS.415.2876B,2012arXiv1203.6641R}.    

Since the linear theory description starts to be valid only at very large scales, and lacking a complete model for general non linear cosmic fluctuations, an extension of the theoretical description has been attempted to non linear and quasi-linear scales thanks to empirical methods based on the so-called streaming model \citep{1980lssu.book.....P}, consisting of linear theory and a convolution on the LOS with a velocity distribution. The model was first adopted on small scales and highly non linear regime to describe the fingers-of-God (FOG) elongation along the LOS due to random motions of virialized objects \citep{1972MNRAS.156P...1J}. Also, fitting functions based on simulation results have been used, for instance, in \cite{1999MNRAS.310.1137H}, \cite{2006MNRAS.368...85T} and \cite{2007MNRAS.374..477T}. Recently, it was shown by \cite{2009MNRAS.393..297P} that on quasi-linear scales a streaming model with a Gaussian velocity dispersion is a good general fit to the redshift-space power spectrum. The goodness of the streaming model was also demonstrated lately, for example, in \cite{2008Natur.451..541G}, \cite{2009MNRAS.393.1183C}, \cite{2011MNRAS.415.2876B}, \cite{2012MNRAS.423.3430B}, \cite{2012arXiv1203.1002M} and \cite{2011arXiv1102.2251C}. We caution that the model breaks down at small scales. This happens even for an unbiased dark matter model (see, for example, \cite{2010PhRvD..82f3522T}) and complex galaxy bias issues maybe required to be under strict control (like shown in \cite{2011ApJ...726....5O,2011MNRAS.417.1913R}). Also, how modes representing large scales are sampled in simulations finite volume boxes might play a role \citep{2008PhRvD..77f3530M,2009PhRvD..80l3503T}, especially if their effect is of the order of a few percent and in the presence of bias, that might further reduce sampling efficiency. Nonetheless, being aware that there exist in the literature works that claim either a good or not so good performance of the streaming model, and that anyway often use it as benchmark, due to its theoretical simplicity and appeal, in this work we will rely on a generalized version of it, with the crucial limitation to deal with appropriately large scales. 

In this paper we find a novel way to measure the $\beta$ parameter on quasi-linear scales in configuration space without utilizing a theoretical description of the linear correlation functions. We generalize a similar result in Fourier space obtained in \cite{1995MNRAS.275..515C}. Our method does not depend on the galaxies mean density and does not employ deconvolutions but only convolutions, an advantage in terms of stability. 

We also note the presence of a baryonic feature at around $100\, \mathrm{Mpc}$ in the so-called normalized quadrupole in configuration space. We are not aware\footnote{After this work was submitted, \cite{2012MNRAS.420.2102S} showed a plot where the feature is visible in simulations results. Also, Eyal Kazin(2011) in a private communication sent us and commented on a similar plot.} of other papers mentioning it, apart from a hint from Fourier space provided in \cite{2010PhRvD..82f3522T} and in relation to biasing from density peaks in \cite{2010PhRvD..81b3526D}. Probably, the reasons for this neglect are that the pioneering RSD papers did not contain a baryonic component and subsequent theoretical analyses were concentrated on relatively smaller scales that were better sampled by the surveys being studied.

The plan of our work is as follows.  We start in Section \ref{sec:theormod} with a review of the theory of RSD in Fourier and configuration space in the linear regime and we discuss the streaming model extension. In Section \ref{sec:methods} we describe our method and the N-body simulations that we use to test it. Section \ref{sec:results} demonstrates how stringently the parameter $\beta$ can be measured, thanks to a full statistical analysis with the simulations and a Fisher matrix study. We conclude in Section \ref{sec:concl}. 

\section{Theoretical Background} \label{sec:theormod}

In what follows we provide a theoretical background both in Fourier and configuration space. It is convenient to be able to switch between the two points of view because, although they are theoretically equivalent, when dealing with real data, they are confronted in practice by somewhat different systematics. Fourier space has an advantage regarding the theoretical treatment because in linear theory the Fourier modes are independent. However, it has been noted that perturbation theory seems to be more effective in configuration space \citep{2008MNRAS.390.1470S,2008PhRvD..77f3530M,2009PhRvD..80l3503T}, i.e. that convergence happens to be faster with respect to Fourier space.

We want to stress that in this work the Fourier treatment is utilized only to derive convenient expressions in configuration space. The observational inputs of our method and manipulations upon them are just in configuration space.  
 
\subsection{Theory in Fourier Space} \label{sec:model2point}

It was shown by \cite{1987MNRAS.227....1K} that in the large-scale linear regime and in the
distant-observer approximation the coherent gravitational infall 
modifies the redshift-space power spectrum in Fourier space:
\begin{equation}\label{eq:kaiserpoint} P_s(k,\mu_k) = (1 + \beta\mu_k^2)^2 P(k),
\end{equation}
where $P(k)$ is the real space power spectrum of the particles that trace the density fluctuations
$\delta$, $\mu_k$ is the cosine of the angle between the wavevector and the LOS,
the subscript $s$ indicates redshift space, and $\beta$ is related to the perturbations growth rate in linear theory.
We define the linear bias $b$ as the ratio of the tracer overdensity $\delta$ with the matter density $\delta_{m}$, \begin{equation}
\label{eq:bias} b \equiv \frac{\delta}{\delta_m}. 
\end{equation} 
The relation between the matter and tracer power spectrum is simply
\be
P(k)=b^2P_m(k).
\ee
We will assume for simplicity a constant bias since we are dealing with large scales and will discuss later the possibility of eventual departures. 
The crucial parameter
$\beta$ is 
\begin{equation} 
\label{fb} \beta \equiv {f(\Omega_m) \over b} = {1 \over b} 
\frac{d\;\ln D}{d\;\ln a} 
\end{equation} 
in which $a$ is the cosmic scale factor and $D$ is the linear density growth
factor. The variable $f$ can be well approximated by
 \begin{equation}
\label{eq:fgamma} f(\Omega_m)=\Omega_m(a)^\gamma 
\end{equation} with $\gamma\simeq0.55$ if general relativity is assumed to be valid \citep{2005PhRvD..72d3529L}; $\Omega_m(a)$ is the fractional density of matter:
\begin{equation}\label{eq:omegaa}
\Omega_m(a)=\frac{H_0^2\Omega_{0m}a^{-3}}{H^2(a)} \end{equation}
where $\Omega_{0m}$ is the matter density today (the subindex $0$ always indicates actual quantities) and
\begin{equation} H(a)=H_0\sqrt{\Omega_{0m}a^{-3}+\Omega_{k}a^{-2}+\Omega_{0\mathrm{de}}a^{-3(1+w)}}
\end{equation}
is the Friedman equation that defines the Hubble parameter $H(a)$ in which $\Omega_{0\mathrm{de}}$ and $w$ are dark energy density parameter and equation of state, and $\Omega_k$ is the curvature energy density.

The linear power spectrum can be decomposed in a multipole expansion \citep{1994MNRAS.267..785C}
\be
P_s(k,\mu_k)= P_0(k)\mathcal{P}_0(\mu_k) + P_2(s)\mathcal{P}_2(\mu_k) +P_4(s)\mathcal{P}_4(\mu_k),
\ee
 in which the multipole coefficients are expressed as
\be
P_l(k)=\frac{2l+1}{2}\int_{-1}^{1} d\mu_k \, \mathcal{P}_l(\mu_k) \, P_s(k,\mu_k),
\ee
where $\mathcal{P}_l(\mu_k)$ is a Legendre polynomial. Each of the coefficients is given in terms of the real space tracer power spectrum
\begin{equation} P_0(k) = \left(1
+ \frac{2\beta}{3} + \frac{\beta^2}{5}\right)P(k), \end{equation}
\begin{equation} P_2(k) = \left(\frac{4\beta}{3} +
\frac{4\beta^2}{7}\right)P(k), \end{equation}
\begin{equation} P_4(k) = \frac{8\beta^2}{35}P(k).
\end{equation} 
Now, if the ratio of the quadrupole to the monopole is taken, this is expressed by a rational function of the $\beta$ parameter \citep{1994MNRAS.267..785C}:
\be
Q(k)=\frac{P_2(k)}{P_0(k)}=\frac{\frac{4}{3}\beta + \frac{4}{7}\beta^2}{1 + \frac{2}{3}\beta +
\frac{1}{5}\beta^2}.
\ee
 Note that, as far as linear theory is valid, this ratio, that we call normalized quadrupole, is scale independent.

\subsection{Theory in Configuration Space}

It is possible to transfer Kaiser's treatment from Fourier to configuration space \citep{1992ApJ...385L...5H},
\begin{equation} \hat{\xi}(\sigma, \pi) = \xi_0(s)\mathcal{P}_0(\mu) + \xi_2(s)\mathcal{P}_2(\mu) +
\xi_4(s)\mathcal{P}_4(\mu), \end{equation} in which $\pi$ is the distance separation along the LOS and $\sigma$ is the perpendicular separation. The absolute distance of
separation is indicated by $s=\sqrt{\sigma^2+\pi^2}$, $\mu=\pi/s$ is the cosine of the angle
between the separation vector and the LOS.

The multipoles of $\hat{\xi}(\sigma, \pi)$ \footnote{\cite{2012MNRAS.419.3223K} have shown an alternative clustering wedges decomposition.} are given in terms of the real space correlation function $\xi(r)$: 
\begin{equation} \xi_0(s) = \left(1
+ \frac{2\beta}{3} + \frac{\beta^2}{5}\right)\xi(r), \end{equation}
\begin{equation} \xi_2(s) = \left(\frac{4\beta}{3} +
\frac{4\beta^2}{7}\right)[\xi(r)-\overline{\xi}(r)], \end{equation}
\begin{equation} \xi_4(s) = \frac{8\beta^2}{35}\left[\xi(r) +
\frac{5}{2}\overline{\xi}(r) -\frac{7}{2}\overline{\overline{\xi}}(r)\right],
\end{equation} 
and 
\begin{equation} \label{eq:obar} \overline{\xi}(r) =
\frac{3}{r^3}\int^r_0\xi(r')r'{^2}dr', \end{equation} \begin{equation} \label{eq:dobar}
\overline{\overline{\xi}}(r) = \frac{5}{r^5}\int^r_0\xi(r')r'{^4}dr'.
\end{equation}

If the real space correlation function $\xi(r)$ is computed from linear theory, these equations describe the typical squashing of the correlation function along the LOS of the Kaiser effect. 
The multipoles of $\xi(\sigma,\pi)$ can be extracted through the projections 
\begin{equation}\label{eq:moment} \xi_{l}(s) =
\frac{2l+1}{2}\int^{+1}_{-1}\xi(\sigma,\pi) \mathcal{P}_{l}(\mu)d\mu. \end{equation}

\cite{1992ApJ...385L...5H} showed that the normalized quadrupole, defined as
\begin{equation}\label{eq:quadru}
Q(s)=\dfrac{\xi_2(s)}{(3/s^3)\int_0^s{\xi_0(s')s'^{2} ds'}-\xi_0(s)},
\end{equation}
can be used to estimate the $\beta$ parameter, since the Kaiser result, valid in linear theory and large scales, brings to the constant ratio
\begin{equation}\label{eq:quadrub} Q(s) =
\frac{\frac{4}{3}\beta + \frac{4}{7}\beta^2}{1 + \frac{2}{3}\beta +
\frac{1}{5}\beta^2}. \end{equation}
All these equations mimic the Fourier description, in particular in both cases the normalized quadrupole on the large scales described by linear theory is a constant plateau.

The multipole correlation functions are related to the corresponding power spectra by
\be \label{eq:ide1}
\xi_l(s)=i^l \frac{1}{2\pi^2}\int_0^{\infty} dk\, k^2\, j_l(ks) \, P_l(k),
\ee
here $j_l$ indicates the spherical Bessel function of the $l$th order
and the following identities presented in \cite{1994MNRAS.267..785C} provide useful connections, that will prove helpful also later on,
\be \label{eq:ide2}
\xi_0(s)-\overline{\xi_0}(s)=-\frac{1}{2\pi^2}\int_0^{\infty} dk\, k^2\, j_2(ks) \, P_0(k)
\ee
\be \label{eq:ide3}
\xi_0(s)+\frac{5}{2}\overline{\xi_0}(s)-\frac{7}{2}\overline{\overline{\xi_0}}(s)=\frac{1}{2\pi^2}\int_0^{\infty} dk\, k^2\, j_4(ks) \, P_0(k).
\ee

\subsection{Departures from Linear Theory}

On very small scales the random motions of the galaxies will produce the FOG phenomenon: an elongation of the correlation function mostly on the LOS direction. 
The FOG effect can be mimicked by a convolution of the correlation function $\hat{\xi}(\sigma, \pi)$ with the distribution function either of random pairwise velocities to give \citep{1980lssu.book.....P}:
\begin{equation} \label{eq:hamiltonmethod} 
\xi(\sigma,\pi) =
\int^{\infty}_{-\infty}\hat{\xi}(\sigma, \pi - v)f(v)dv \end{equation}
where the peculiar velocities are expressed in comoving
coordinates. The convolution device is usually defined as the streaming model.

The random motions of particles belonging to a virialized system depend on how the galaxies are selected and have been represented in the literature by an exponential form, \begin{equation} f(v) =
\frac{1}{\sigma_{v}\sqrt{2}}\exp\left(-\frac{\sqrt{2}|v|}{\sigma_{v}}\right)
\label{e:fv} \end{equation} or a Gaussian form \begin{equation} f(v) =
\frac{1}{\sqrt{2\pi\sigma^2_{v} }}\exp\left(-\frac{v^2}{2\sigma^2_{v}}\right)
\label{e:fvgauss} \end{equation}
where $\sigma_{v}$ is the pairwise peculiar velocity dispersion. The exponential and Gaussian
forms have proved to be good fits of the observed data and simulations  \citep[see, for example,][]{1996MNRAS.280L..19P,1998ASSL..231..185H,2001Natur.410..169P,2008Natur.451..541G,2009MNRAS.393.1183C,2011MNRAS.415.2876B,2012MNRAS.423.3430B,2012arXiv1203.1002M,2011arXiv1102.2251C}. In this work our main concern is on relatively large scales and our results do not depend much on the choice of the distribution function. 
An important addition to the applicability of the above convolution has been offered by \cite{2009MNRAS.393..297P}. They showed that the convolution with $\sigma_v$ as a free parameter could be useful to describe \emph{quasi-linear} scales important for the transition to linear scales. As we mentioned in the Introduction, at small enough scales non linearities and galaxy bias will start to spoil the simplicity of the model. How small these limit scales have to be depends also on the bias of the set of galaxies under study.

Carrying over the convolution to Fourier space, one simply gets an additional factor to the Kaiser formulation, and what we define as the observed power spectrum looks like
\be \label{eq:modelPW}
P^{\textrm{obs}}(k,\mu_k)= (1+\beta\mu_k^2)^2\, P(k) \,\hat{f}(\sigma_v k\mu_k),
\ee
where $\hat{f}(\sigma_v k\mu_k)$ is the Fourier transform of the velocity distribution. If the distribution is a Gaussian, it follows that
\be
\hat{f}(\sigma_v k \mu_k)=e^{-(\sigma_v k \mu_k)^2/2},
\ee
while if it is an exponential we have that
\be
\hat{f}(\sigma_v k \mu_k)=\frac{1}{1+(\sigma_v k \mu_k)^2/2}.
\ee
Both forms have the same Taylor expansion for a small $\sigma_v$ up to second order.
The projected multipoles of the observed power spectrum, that we also refer to as the observed multipoles, can be found from
\be
P^{\textrm{obs}}_l(k)=\frac{2l+1}{2}\int_{-1}^{1} d\mu_k \, \mathcal{P}_l(\mu_k) \, P^{\textrm{obs}}(k,\mu_k).
\ee

Eventual random errors on redshift determinations can be described by a convolution with a similar Gaussian.
Given the literature results favorable to the streaming model, we assume that as long as we refer to scales larger than the effective $\sigma_v$ parameter, all the above mentioned deviations from linear theory with a linear bias can be reasonably controlled by the streaming model with the free parameter $\sigma_v$. Clearly, if one is able to predict the dispersion parameter from first principles, all the attention could then be concentrated on the $\beta$ parameter, that could be determined with more precision.

\section {A new method to estimate the redshift-space distortions parameter} \label{sec:methods}

We now introduce our new method starting from the problems that originate from the normalized quadrupole defined in Equation\,(\ref{eq:quadru}). 
 We test our method with the results from the $N$-body simulations by \cite{2011PhRvD..84d3501S} to estimate the level of systematic and statistical errors for the recovery of the parameter $\beta$. We also perform a Fisher matrix analysis to check on the expected error magnitude.

As in the Sato and Matsubara simulations, our chosen cosmological parameters are: $\Omega_{\Lambda}=0.735$, $\Omega_m=1-\Omega_{\Lambda}$, $\Omega_b=0.0448$, $w=-1$, $h=0.71$, $n_s=0.963$, and $\sigma_8=0.8$. These are, respectively, the fractional density of dark energy, dark matter and baryons, the dark energy equation of state, the Hubble parameter (in units of $100\,\mathrm{km\,s^{-1}\,Mpc^{-1}}$), the primordial perturbations spectral index and the normalization factor relative to the variance of the density fluctuations contained within a scale of $8\,h^{-1}\,\mathrm{Mpc}$. The 30 simulations that we utilize were done with the {\em Gadget2} \citep{2005MNRAS.364.1105S} code with $N_p=1024^3$ dark matter particles, a box size $L_{\mathrm{box}}=1000\,h^{-1}\,\mathrm{Mpc}$, a softening length $r_s=50\,h^{-1}\,\mathrm{kpc}$, an initial redshift $z_{\mathrm{ini}}=36$, an output redshift $z=1$. 

As mentioned by the authors there might be a slight systematic effect due to finite-mode sampling in a limited number of finte-size boxes. To quantify such an uncertainty we set up for an approximate correction to the data. As suggested in \cite{2011PhRvD..84d3501S} and \cite{2009PhRvD..80l3503T}, we correct for finite-mode sampling by using simulations at a redshift when linear theory dominates. The simulations at the initial redshift $z_{\mathrm{ini}}=36$, also kindly provided to us by Sato and Matsubara, are ideal for our scope. We proceed by rescaling the $z_{\mathrm{ini}}=36$ monopoles and quadrupoles to $z=1$ using linear theory. We then subtract those extrapolated monopoles and quadrupoles from the original $z=1$ snapshots. We thus obtain at this point a representation of only the non linear contributions due to the fact that the rescaled functions have the same initial random seeds as the original $z=1$ snapshots. As a final step, we generate random linear monopoles and quadrupoles, properly correlated among each other, and we add them to the above non linear-only parts. We have checked, using the untouched $z=1$ simulations, that the final statistical errors on $\beta$ do not change significantly, therefore justifying our approximate correction. We find that the extraction of the $\beta$ parameter is corrected from a bias of about 4\%--5\% with respect to carrying out a parallel analysis on the uncorrected data. This might be due to the fact that the quadrupole, that is very sensitive to the $\beta$ parameter, is quite susceptible to the finite-mode sampling problem, since it is, for example, less sampled than the monopole. Since we are expecting statistical errors on $\beta$ at the few percent level, as the simulations and Fisher matrix analyses exposed later show, we employ our corrections in the rest of the paper.

\begin{figure}[!t] 
\centering
\includegraphics[scale=0.58]{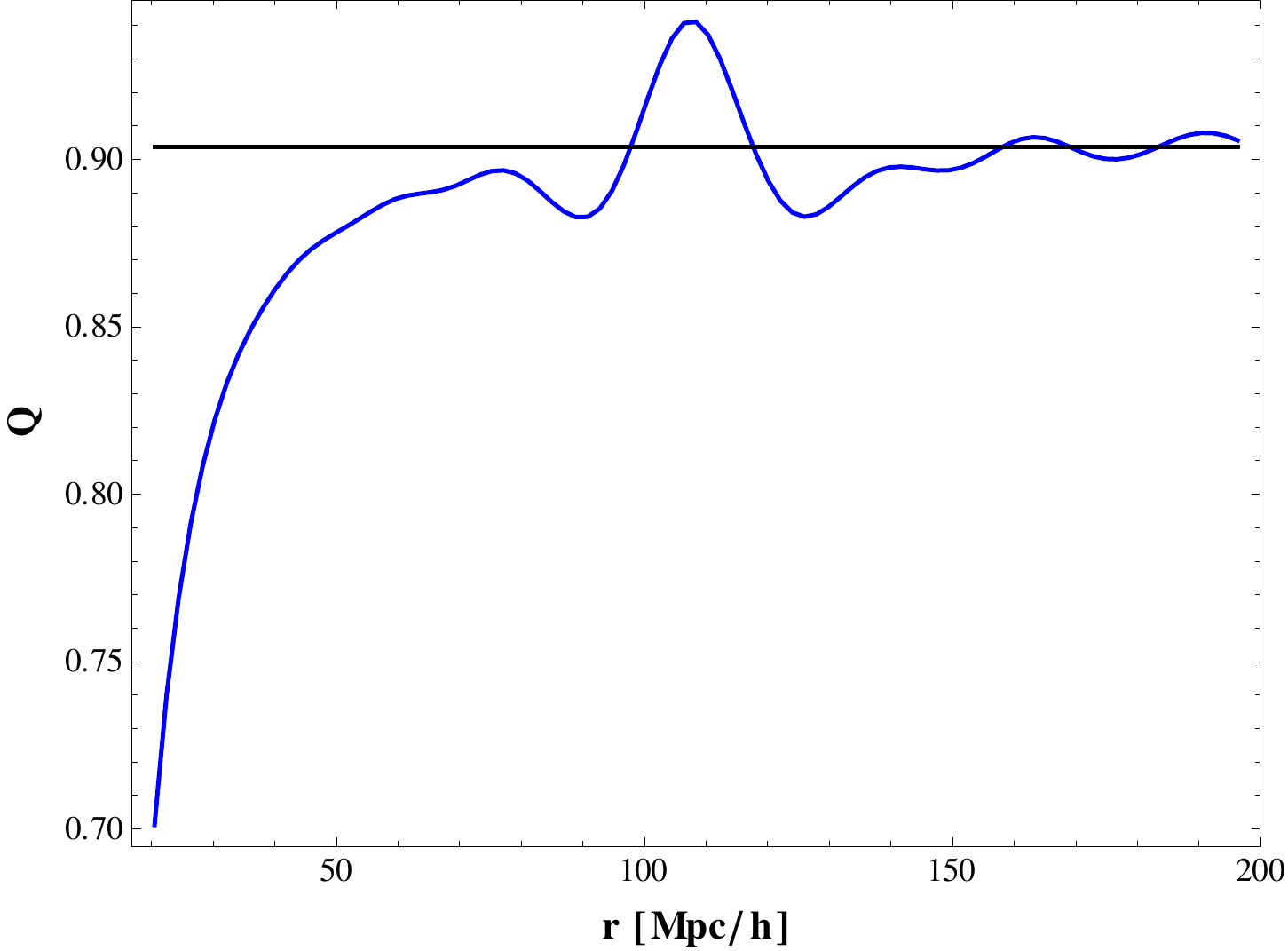}
\caption{ \label{fig:Qlin}
Normalized quadrupole (dashed line) defined in Equation\,(\ref{eq:quadru}) for the streaming model. The linear theory correlation function is convolved along the line of sight with a Gaussian velocity distribution.  The asymptote (solid line) Equation\,(\ref{eq:quadrub}) is reached at very large scales and furthermore the reading of the plateau around $100\,h^{-1}\mathrm{Mpc}$ is complicated by the presence of residuals of baryonic acoustic oscillations.
}
\end{figure}

In Figure\,(\ref{fig:Qlin}) we show the normalized quadrupole derived using the linear power spectrum computed by $CAMB$ \citep{2000ApJ...538..473L} for the fiducial cosmology, that implies for dark matter $\beta=0.85$, and a Gaussian velocity dispersion with $\sigma_v=5.7\,h^{-1}\,\mathrm{Mpc}$. It can be noticed that the asymptotic constant form is reached only at very large distances, where the error due to cosmic variance is supposed to be important, and residual baryonic acoustic oscillations at scales around $100\,h^{-1}\,\mathrm{Mpc}$ can potentially confuse the estimation of the plateau. \cite{2010PhRvD..82f3522T} find oscillations in the Fourier space ratio of the quadrupole and monopole power spectra. We speculate that the baryonic spike we observed using a simple phenomenological streaming model will be somewhat enhanced due to the extra terms found in \cite{2010PhRvD..82f3522T} for unbiased dark matter.
The feature should perturb the extraction of the growth parameter in configuration space from the asymptotic limit of the normalized quadrupole if scales close to its location are employed.

One could proceed by comparing the observed normalized quadrupole with the theoretical estimated one. However this would be done at the cost of introducing a dependence on the cosmological parameters, which would have to be marginalized over to obtain a statistically independent error on $\beta$. The above marginalizations are going to inflate the error on $\beta$. If not predicted properly, the dispersion $\sigma_v$ has to be marginalized over.

We propose here a new method to estimate the $\beta$ parameter that depends just on the observed monopole and quadrupole components of the observed correlation function and not on their linear theory representation, including its dependence on cosmological parameters.

To derive our estimator we start in Fourier space to generate as a final product a configuration space expression. The observational inputs and their further handling belong all to configuration space.
Our starting point is Equation\,(\ref{eq:modelPW}), that was shown to be a plausible convenient description on scales characterized by quasi-linear effects in \cite{2009MNRAS.393..297P}. We also mention that such a model is a worth starting point, since it is practically always considered in the literature as a benchmark against which more sophisticated models are confronted with. Furthermore, quantities in configuration space have been shown to be described by perturbation theory somewhat better than the Fourier space counterparts.

The observed monopole and quadrupole can be written in terms of the linear spectra:
\bea
P_0^{\textrm{obs}}(k)&=&P(k)\frac{1}{2}\int_{-1}^{1} d\mu_k \, \mathcal{P}_0(\mu_k) \, (1+\beta \mu_k^2)^2\, \hat{f}(\sigma_v k\mu_k) \nonumber\\ 
&=&  h_0(\beta,\sigma_v k)  P(k)
\eea
and
\bea
P_2^{\textrm{obs}}(k)&=&P(k)\frac{5}{2}\int_{-1}^{1} d\mu_k \, \mathcal{P}_2(\mu_k) \, (1+\beta \mu_k^2)^2\, \hat{f}(\sigma_v k\mu_k) \nonumber\\
&=&  h_2(\beta,\sigma_v k)  P(k)
\eea
The functions $h$ are
\bea
h_0(\beta,\sigma_v k) &=&\left(1
+ \frac{2\beta}{3} + \frac{\beta^2}{5}\right)f_{00}(k\sigma_v) \\
&+&\left(\frac{4\beta}{3} +
\frac{4\beta^2}{7}\right)f_{02}(k\sigma_v)+\frac{8\beta^2}{35}f_{04}(k\sigma_v) \nonumber
\eea
and
\bea
h_2(\beta,\sigma_v k) &=&\left(1
+ \frac{2\beta}{3} + \frac{\beta^2}{5}\right)f_{20}(k\sigma_v) \\
&+&\left(\frac{4\beta}{3} +
\frac{4\beta^2}{7}\right)f_{22}(k\sigma_v)+\frac{8\beta^2}{35}f_{24}(k\sigma_v). \nonumber
\eea
The functions $f$ are given by
\be
f_{0l}(k\sigma_v)=\frac{1}{2}\int_{-1}^{1} d\mu_k \, \hat{f}(\sigma_v k\mu_k) \, \mathcal{P}_0(\mu_k)  \mathcal{P}_l(\mu_k)
\ee 
and
\be
f_{2l}(k\sigma_v)=\frac{5}{2}\int_{-1}^{1} d\mu_k \, \hat{f}(\sigma_v k\mu_k) \, \mathcal{P}_2(\mu_k)  \mathcal{P}_l(\mu_k),
\ee
\\
where $l=0,2,4$. These functions can be expressed analytically in terms of the error or gamma functions for a Gaussian distribution or in terms of the $\arctan$ function for an exponential distribution \citep{1995MNRAS.275..515C,1996MNRAS.280L..19P}.

It was shown in \cite{1995MNRAS.275..515C} that, in the regime where Equation\,(\ref{eq:modelPW}) applies, the ratio
\be \label{eq:ratioColes}
\frac{P_2^{\textrm{obs}}(k)}{P_0^{\textrm{obs}}(k)}=\frac{h_2(\beta,\sigma_v k)}{h_0(\beta,\sigma_v k)}
\ee
does not depend on the linear power spectrum $P(k)$ and therefore can be used on large scales to determine the $\beta$ parameter without the need to marginalize on the cosmological parameters. We point out here that the ratio in Equation\,(\ref{eq:ratioColes}) induces also the disappearance of \textit{any} additional function of wavenumber $k$ multiplying the power spectrum, possibly approximating non linear corrections and/or a scale dependent galaxy bias. Even better, the method does not require the knowledge of functions of just $k$ in the $P^{\textrm{obs}}(k,\mu_k)$ expression, and it is therefore more general than the streaming model, where $P^{\textrm{obs}}(k,\mu_k)$ has to be fully specified. In this work we present a new way to exploit these nice properties in configuration space, where there might be more favorable convergence properties in perturbation theory respect to Fourier space. The identities
\be \label{eq:defDk}
\mathcal{D}(k)\equiv h_0(\beta,\sigma_v k)P_2^{\textrm{obs}}(k)-h_2(\beta,\sigma_v k)P_0^{\textrm{obs}}(k)=0
\ee
and
\bea \label{eq:defD}
D(s)&\equiv&-\frac{1}{2\pi^2}\int_0^{\infty} dk\, k^2\, j_2(ks) \,h_0(\beta,\sigma_v k)P_2^{\textrm{obs}}(k) \nonumber\\
&+&\frac{1}{2\pi^2}\int_0^{\infty} dk\, k^2\, j_2(ks) \,h_2(\beta,\sigma_v k)P_0^{\textrm{obs}}(k) \nonumber\\
&\equiv& D_2(s)-D_0(s)
\eea
conveniently summarize the basis for our method. We first manipulate the ratio in Equation\,(\ref{eq:ratioColes}) to reduce it to the zero quantity in Fourier space $\mathcal{D}$, and then we apply the operator $-\frac{1}{2\pi^2}\int_0^{\infty} dk\, k^2\, j_2(ks)$, with the newly formed difference $D$, that is a configuration space function,  still being equal to zero. The transformed multipoles are labeled by $D_0$ and $D_2$. We were then able to express the integrals in terms of solely $\xi^{\textrm{obs}}_0$, $\xi^{\textrm{obs}}_2$, objects in configuration space, and $\sigma_v$ and $\beta$, without any dependence on the theoretical linear correlation functions. The dependence on the cosmological parameters is therefore concentrated exclusively on $\beta$. The final result is
\begin{widetext}
\bea \label{eq:estD}
D&=&b_0\left( C_{00} \left[ \xi^{\textrm{obs}}_2-\tilde{\xi}^{\textrm{obs}}_2 \right] - \overline{C_{00} \left[ \xi^{\textrm{obs}}_2-\tilde{\xi}^{\textrm{obs}}_2 \right]}\right)+ b_2\left( C_{02} \left[ \xi^{\textrm{obs}}_2 \right] - \overline{C_{02} \left[ \xi^{\textrm{obs}}_2 \right]}\right) \nonumber\\
&+&b_4\left( C_{04} \left[ \xi^{\textrm{obs}}_2-\frac{7}{5} \overline{\overline{\xi}}^{\textrm{obs}}_2 \right] - \overline{C_{04} \left[ \xi^{\textrm{obs}}_2-\frac{7}{5} \overline{\overline{\xi}}^{\textrm{obs}}_2 \right]}\right) \nonumber\\
&-&b_0\left( C_{20} \left[ \xi^{\textrm{obs}}_0 \right] \right)-b_2\left(  C_{22} \left[ \xi^{\textrm{obs}}_0 - \overline{\xi}^{\textrm{obs}}_0 \right] \right) -b_4\left(  C_{24} \left[ \xi^{\textrm{obs}}_0 +\frac{5}{2} \overline{\xi}^{\textrm{obs}}_0-\frac{7}{2} \overline{\overline{\xi}}^{\textrm{obs}}_0 \right] \right).
\eea
\end{widetext}
We now explain all the terms appearing in the difference $D$, that is in general a function of separation distance, but is equal to zero when the model given by Equation\,(\ref{eq:modelPW}), even thought as multiplied by a generic function of $k$, is valid, the true parameters $\beta$ and $\sigma_v$ are used and in the absence of noise. We expect that a strong dependence on scale will show up at small scales due to highly non linear behavior. The single and double overline were defined in Equations\,(\ref{eq:obar}) and (\ref{eq:dobar}). The constants $b$ are
\bea
b_0&=&\left(1+\frac{2\beta}{3}+\frac{\beta^2}{5}\right) \nonumber\\
b_2&=&\left(\frac{4\beta}{3} +\frac{4\beta^2}{7}\right) \nonumber\\
b_4&=&\frac{8\beta^2}{35}.
\eea
The convolutions $C$ depend on the distribution $f(v)$ and on its variance $\sigma_v$. In configuration space, when applied on a generic function $g(s)$, they are
\bea
C_{0l}\left[g\right](s)&\equiv&\frac{1}{2}\int^{+1}_{-1}d\mu \mathcal{P}_{0}(\mu) \int^{\infty}_{-\infty}dv f(v) \mathcal{P}_l(\mu') g(s')\nonumber\\
C_{2l}\left[g\right](s)&\equiv&\frac{5}{2}\int^{+1}_{-1}d\mu \mathcal{P}_{2}(\mu) \int^{\infty}_{-\infty}dv f(v) \mathcal{P}_l(\mu') g(s'),\nonumber\\
&&
\eea
where we used
\bea
s'&\equiv&\sqrt{s^2+v^2-2\mu sv} \nonumber \\
\mu'&\equiv&\frac{\mu s-v}{s'}
\eea
and $l=0,2,4$.
The terms with the convolution operators $C$ in Equation\,(\ref{eq:estD}) are given by
\begin{align} \label{eq:eq00}
& C_{00} \left[ \xi^{\textrm{obs}}_2-\tilde{\xi}^{\textrm{obs}}_2 \right](s) - \overline{C_{00} \left[ \xi^{\textrm{obs}}_2-\tilde{\xi}^{\textrm{obs}}_2 \right]}(s)=\\
& -\frac{1}{2\pi^2}\int_0^{\infty} dk\, k^2\, j_2(ks) \, P^{\textrm{obs}}_2(k)\, f_{00}(k\sigma_v), \nonumber
\end{align}
\begin{align}
& C_{02} \left[ \xi^{\textrm{obs}}_2 \right](s) - \overline{C_{02} \left[ \xi^{\textrm{obs}}_2 \right]}(s)=\\
& -\frac{1}{2\pi^2}\int_0^{\infty} dk\, k^2\, j_2(ks) \, P^{\textrm{obs}}_2(k)\, f_{02}(k\sigma_v), \nonumber
\end{align}
\begin{align} \label{eq:eq04}
&  C_{04} \left[ \xi^{\textrm{obs}}_2-\frac{7}{5} \overline{\overline{\xi}}^{\textrm{obs}}_2 \right](s) - \overline{C_{04} \left[ \xi^{\textrm{obs}}_2-\frac{7}{5} \overline{\overline{\xi}}^{\textrm{obs}}_2 \right]}(s)=\\
&  -\frac{1}{2\pi^2}\int_0^{\infty} dk\, k^2\, j_2(ks) \, P^{\textrm{obs}}_2(k)\, f_{04}(k\sigma_v), \nonumber
\end{align}
and
\bea
&&C_{20} \left[ \xi^{\textrm{obs}}_0 \right](s) = \\
&{}& -\frac{1}{2\pi^2}\int_0^{\infty} dk\, k^2\, j_2(ks) \, P^{\textrm{obs}}_0(k)\, f_{20}(k\sigma_v), \nonumber
\eea
\bea
&&C_{22} \left[ \xi^{\textrm{obs}}_0 - \overline{\xi}^{\textrm{obs}}_0 \right](s)= \\
&{}&- \frac{1}{2\pi^2}\int_0^{\infty} dk\, k^2\, j_2(ks) \, P^{\textrm{obs}}_0(k)\, f_{22}(k\sigma_v), \nonumber
\eea
\bea
&&C_{24} \left[ \xi^{\textrm{obs}}_0 +\frac{5}{2} \overline{\xi}^{\textrm{obs}}_0-\frac{7}{2} \overline{\overline{\xi}}^{\textrm{obs}}_0 \right](s)= \\
&{}&- \frac{1}{2\pi^2}\int_0^{\infty} dk\, k^2\, j_2(ks) \, P^{\textrm{obs}}_0(k)\, f_{24}(k\sigma_v). \nonumber
\eea
The formulae in Equations\,(\ref{eq:ide1})--(\ref{eq:ide3}) were used to obtain the above identities. Furthermore, in Equations\,(\ref{eq:estD}) and (\ref{eq:eq00}) we employed the definition
\bea
\tilde{\xi}^{\textrm{obs}}_2(r)\equiv3\int_{r}^{\infty} dr' \, \frac{\xi^{\textrm{obs}}_2(r')}{r'}
\eea
together with the identity
\bea
 j_0(kr)+ j_2(kr)=3\int_{r}^{\infty} dr'  \, \frac{j_2(kr')}{r'}=\tilde{j}_2(kr),
\eea
while in Equations\,(\ref{eq:estD}) and (\ref{eq:eq04}) we utilized Equation\,(\ref{eq:dobar}),
\bea
\overline{\overline{\xi}}^{\textrm{obs}}_2 = \frac{5}{r^5}\int^r_0 \xi^{\textrm{obs}}_2(r')r'{^4}dr',
\eea
in conjunction with
\bea
 j_2(kr)+ j_4(kr)=\frac{7}{r^5}\int^r_0 j_2(kr') r'{^4}dr'=\frac{7}{5} \overline{\overline{j}}_2(kr).
\eea

In practical terms it is easy to build an efficient routine to compute the estimator in Equation\,(\ref{eq:estD}). The dependence on $\beta$ is analytical and only a one-dimensional grid on $\sigma_v$ is necessary. A single case with the two parameters fixed took us a few seconds on a desktop workstation.

When ideally applied on the observed correlation functions from the model of Equation\,(\ref{eq:modelPW}), linear theory convolved with a velocity dispersion, the difference estimator $D$ completely undoes the convolution with the velocity dispersion and returns a zero response if using the true $\beta$ and $\sigma_v$. This is expected since $D$ was explicitly based on the model in Equation\,(\ref{eq:modelPW}). 
\begin{figure}[!t] 
\centering
\includegraphics[scale=0.65]{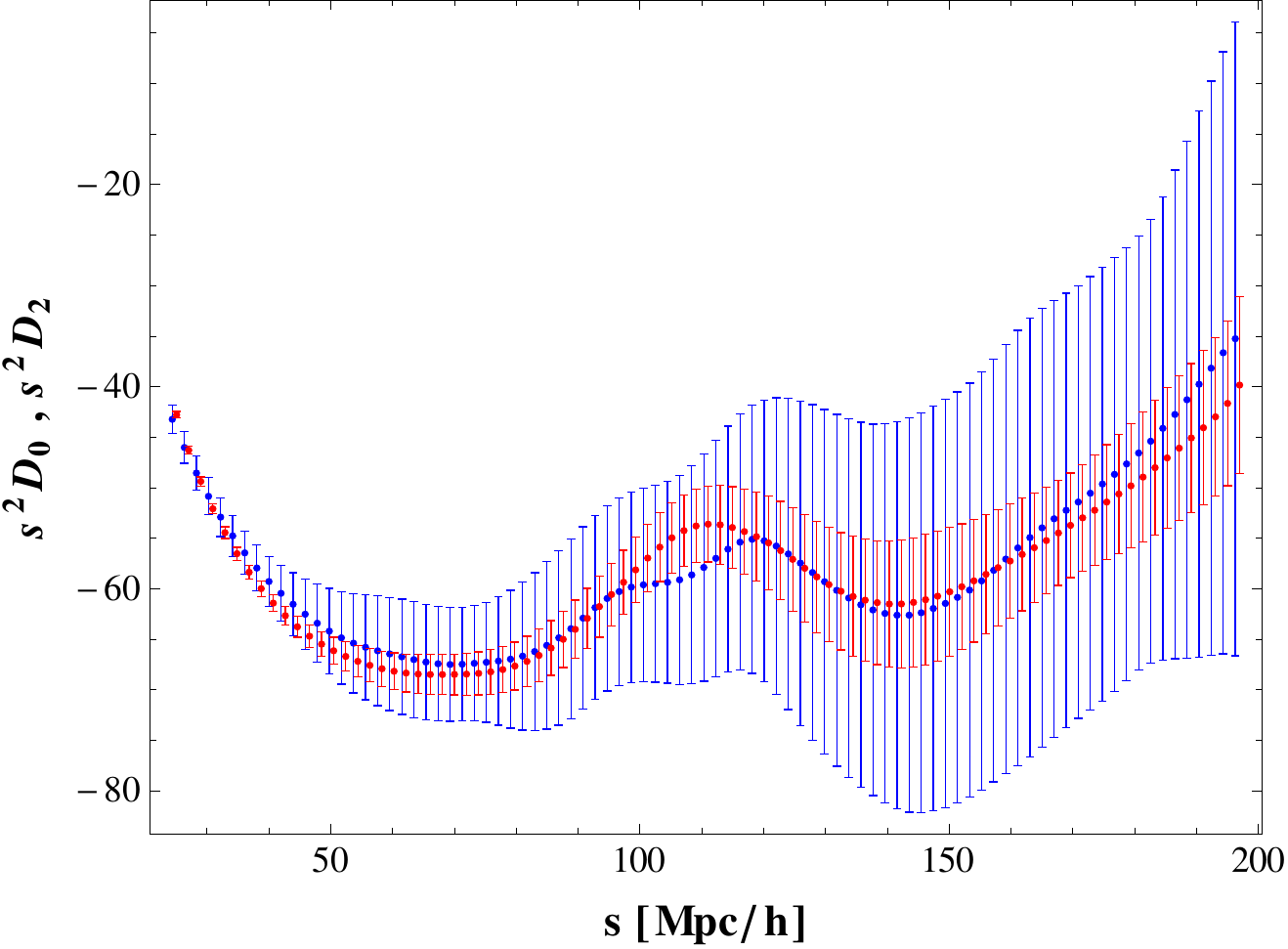}
\caption{\label{fig:D}
Effect of the difference estimator $D$ given by Equations\,(\ref{eq:defD}) and (\ref{eq:estD}) as seen through the two transformed multipoles terms $D_0$ and $D_2$ subtracting each other in the rhs of Equation\,(\ref{eq:defD}). For better visibility, we have artificially shifted one of the curves by $0.7\,h^{-1}\mathrm{Mpc}$ and we have multiplied them by the square of the separation distance. The correlation functions derived from the Sato and Matsubara $N$-body simulations described in the text are convolved along the line of sight with a Gaussian velocity distribution according to our prescription and we present an average of the different realizations. Since the two curves are very close to each other, the prescription $D$ is capable of undoing most of small non linearities for scales larger than about $30\,h^{-1}\mathrm{Mpc}$. In this example we fixed the parameters to $\beta=0.85$ and $\sigma_v=5.7\,h^{-1}\,\mathrm{Mpc}$. The bulk of the baryonic acoustic oscillations is very similar in the two cases. The disparity in the uncertainties, here obtained from the diagonal elements of the covariance matrix, comes about because the term derived from the monopole has much smaller errors respect to the quadrupole one. The role of $\sigma_v$ consists in trying to cope with mild nonlinearities given the range where the fit is performed. In the full statistical analysis $\sigma_v$ is marginalized over.
}%
\end{figure}
Much more interesting and realistic is when we consider the correlation functions built from the Sato and Matsubara $N$-body simulations. It is important to test if and at which scales
the action of $D$ is able to nullify or markedly reduce the overall effect of non linearities and convolution with the velocity dispersion. An example of the effectiveness of the prescription $D$ from Equations\,(\ref{eq:defD}) and (\ref{eq:estD}) is provided in Figure\,\ref{fig:D}. It can be noticed that the two terms of the right-hand side (rhs) of Equation\,(\ref{eq:defD}), the transformed multipoles $D_0$ and $D_2$, are very similar to each other. Also, the bulk of the baryonic feature is similar for both cases, apart from a small non linear contribution. This means that our proposed processing on the monopole and quadrupole is capable of bringing the final products to a close agreement. The plot is just an example to display the general idea of the method. In the next section, we perform a full $\chi^2$ analysis on the simulation data formed from the total difference $D$ to find if the best-fitting $\beta$ is close to the fiducial value. As already mentioned, we consider only a Gaussian velocity dispersion. But if needed, more effective velocity dispersions might be easily implemented within the same framework in future analyses.

Two nice properties of the statistics $D$ are that it does not depend on the mean density of the galaxies and that it is defined in terms of convolutions. The first attribute can be easily proved by looking at Equation\,(\ref{eq:defD}), where the presence of the Bessel function assures that the zero-mode contribution vanishes. The second feature potentially provides a certain degree of stability, compared with methods that make use of deconvolutions.

To widen the range of applicability, the method will have to be further tested with $N$-body simulations that include also galaxy bias. However, due to its simplicity and practicality, we believe that, with in case necessary phenomenological modifications to account for some of the effects of non linearities and galaxy bias, our method could be a useful tool to study RSDs.

\section {Results} \label{sec:results}
In a practical situation where data from a galaxy survey are available, the quantity $D$ in Equation\,(\ref{eq:estD}) can be used directly in configuration space.
Here the correlation functions can be measured, for example, by the \cite{1993ApJ...412...64L} estimator. We recommend the application of the following $\chi^2$ function in order to select from the data the best $\beta$ and $\sigma_v$. The covariance matrix $\mathrm{Cov}D_{ij}$ relative to the distances $s_i$ and $s_j$ for $D$ could be computed from for example a set of $N$-body simulations, like we are doing in this work, according to
\bea
\mathrm{Cov}D_{ij}&=&\frac{1}{N_{\mathrm{run}}-1} \nonumber \\
&\times& \sum_{\mathrm{irun}=1}^{N_{\mathrm{run}}} \left( D^{\mathrm{irun}}(s_i,\beta^{\mathrm{fid}},\sigma^{\mathrm{fid}}_v)-\bar{D}(s_i,\beta^{\mathrm{fid}},\sigma^{\mathrm{fid}}_v) \right) \nonumber \\
&\times & \left( D^{\mathrm{irun}}(s_j,\beta^{\mathrm{fid}},\sigma^{\mathrm{fid}}_v)-\bar{D}(s_j,\beta^{\mathrm{fid}},\sigma^{\mathrm{fid}}_v) \right),
\eea
where $\mathrm{irun}$ is the realization index, that reaches the total $N_{\mathrm{run}}=30$ in our case. The overbar indicates the mean over the realizations and the parameters $\beta$ and $\sigma_v$ are considered fixed to fiducial values, that need to be relatively close to the best-fit values. A few extra iterations can update these values if necessary.
It is then natural to evaluate  
\be \label{eq:chisq}
\chi^2=\sum_{i,j} D(s_i,\beta,\sigma_v)\,\, \mathrm{Cov}D_{ij}^{-1}\, D(s_j,\beta,\sigma_v),
\ee 
where we have explicitly indicated that $D$ is a function of distance separation, $\beta$ and $\sigma_v$. We apply this statistic to the simulations results.

The monopole and quadrupole functions furnished to us by Sato and Matsubara are evaluated at discrete points separated by $\Delta r=2\,h^{-1}\mathrm{Mpc}$. It turns out that the derived covariance matrix $\mathrm{Cov}D_{ij}$ is highly correlated and close to singular and in order to be able to stably invert it we have used singular value decomposition (SVD). The smaller eigenvalues, responsible for the instability, are more affected by noise than the larger ones and it is necessary to operate a cut upon them, i.e. replace their inverses by zeros in the SVD representation of the inverse covariance matrix. The situation is more conservative for the first eigenvalues but the statistical errors are larger, therefore a balance has to be struck. 

We adopt the following method for the eigenvalue selection. Assuming that the first $p$ eigenvalues surviving the cut are almost clean from noise, after having found the best-fit $\beta$ and $\sigma_v$, the corresponding $\chi^2$ will have effectively $p-2$ degrees of freedom, in the presence of a noise close to Gaussian. Hence we choose to place the cut where $\chi^2_{p-2}$ can no longer be accepted as a null hypothesis. In practical terms, we summed the $\chi^2_{p-2}$ for each of the $N_{\mathrm{run}}=30$ realizations to form a total $\chi^2_{N_{\mathrm{run}}(p-2)}$ statistic. Then we considered the variable $\sqrt{2 \chi^2_{N_{\mathrm{run}}(p-2)}}$, that is empirically known to approximately follow a Gaussian with mean $\sqrt{2 N_{\mathrm{run}}(p-2)-1}$ and unit variance if $N_{\mathrm{run}}(p-2)$ is greater than 30, as it is practically always the case in our computations. We verified that a stable criterion of rejection is when the statistical significance of the test of the correctness of the null hypothesis was smaller than $5\%$. To decide over borderline cases we added a Kolmogorov--Smirnov test with the same rejection barrier of $5\%$. Rejection was conservatively enacted even if only one of the two tests was passed. We found that the two tests gave very similar results and that the estimated $\beta$, found by averaging over the $N_{\mathrm{run}}=30$ best-fit values obtained by minimizing Equation\,(\ref{eq:chisq}) for each simulation realization, and its error, found from the standard deviation of the best fits, were very stable for eigenvalues neighboring the cut. 

\begin{figure}[!t]
\centering
\includegraphics[scale=0.65]{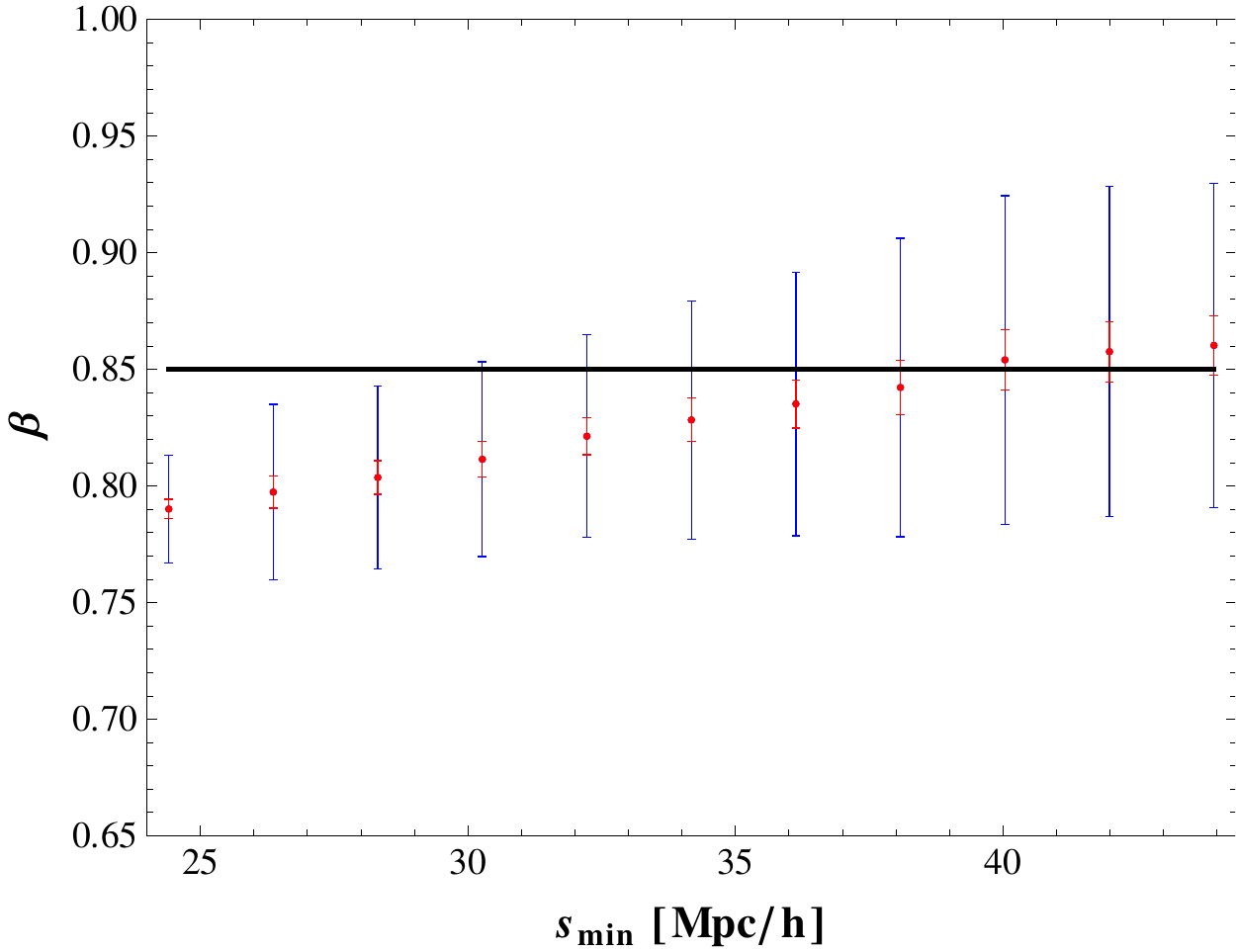}
\caption{\label{fig:datapoints}
Estimations of the $\beta$ parameter from the Sato and Matsubara $N$-body simulations. We show the marginalized $1\sigma$ error bars for ranges of distance separations going from a variable $s_{\mathrm{min}}$ to a fixed $s_{\mathrm{max}}=80\,h^{-1}\mathrm{Mpc}$, after which there is no statistical gain, given the simulations box volume. The data have been obtained by averaging and forming the standard deviations of the marginalized samples produced by a maximum likelihood analysis applied to every realization. The fiducial value depicted by a solid line is compatible with the data from $s_{\mathrm{min}}=30\,h^{-1}\mathrm{Mpc}$. While the data are consistent with no bias from $s_{\mathrm{min}}=35\,h^{-1}\mathrm{Mpc}$, since the errors on the mean, shown as the smaller superimposed error bars, start to be less that $2\sigma$ from the fiducial value.
}
\end{figure}

Concerning the range of separation distances used in the analysis, we fixed the maximum radius to $s_{\mathrm{max}}=80\,h^{-1}\mathrm{Mpc}$, as with our box volume no statistical improvements were evident for larger separations. Instead, we varied the minimal separation, that should crucially indicate to us where systematics due to non linearity are important. The best-fit maximum likelihood values of the parameters constitute marginalized samples. As shown in Figure\,\ref{fig:datapoints}, we found that for a minimum separation such that $s_{\mathrm{min}}>30\,h^{-1}\mathrm{Mpc}$ the retrieved $\beta$ was compatible at $1\sigma$ with the fiducial input value $\beta^{\mathrm{fid}}=0.85$ with a marginalized statistical error of $5\%$. While for separations such that $s_{\mathrm{min}}>35\,h^{-1}\mathrm{Mpc}$, even the errors on the mean were consistent with $\beta^{\mathrm{fid}}$. At $s_{\mathrm{min}}=35\,h^{-1}\mathrm{Mpc}$ we found that the marginalized statistical error on $\beta$ is $6\%$. We conclude that for dark matter boxes of volume $V_s=1\,\left( h^{-1}\,\mathrm{Gpc} \right)^3$, we obtain reliable results for $s_{\mathrm{min}}>30\,h^{-1}\mathrm{Mpc}$, since the statistical errors are large enough to provide for possible systematic ones. While for $s_{\mathrm{min}}>35\,h^{-1}\mathrm{Mpc}$ we do not see evidence for bias and we could imagine to apply our method to larger volumes with smaller statistical uncertainties. 

In order to theoretically estimate the magnitude of the error on the parameter $\beta$ given a galaxy survey configuration, we employ the Fisher matrix formalism applied to Fourier space. This technique allows us to propagate the uncertainties on measured quantities into an uncertainty on $\beta$. We pursue the following schema: we interpret the difference $\mathcal{D}$ as a function depending on the monopole and quadrupole projections of the observed power spectrum, as in Equation\,(\ref{eq:defDk}); we compute the errors on those projections; we propagate those errors on $\mathcal{D}$; finally we transfer the uncertainties on the $\beta$ and $\sigma_v$ parameters. This calculation is based on the model of Equation(\ref{eq:modelPW}), linear theory convolved with a velocity dispersion. We choose the same cosmology used in the simulations.

The variance of the measurement error on the observed power spectrum \citep{1994ApJ...426...23F} is
\be
\left[ \Delta P^{\textrm{obs}}(k,\mu_k) \right]^2=\frac{(2\pi)^2 }{V_s k^2 \Delta k \Delta \mu_k} \left[  P^{\textrm{obs}}(k,\mu_k)+\frac{1}{\overline{n}} \right]^2,
\ee
where $k$ is a specific wavenumber at the center of the interval $\Delta k$, $V_s$ is the survey geometric volume, and $\overline{n}$ is the mean number density of the tracer. 
We can then build the covariance matrix for the multipoles of the observed power spectrum \citep{2006PASJ...58...93Y,2003ApJ...595..577Y}:
\bea
G_{lm}(k)&=&\frac{(2l+1)(2m+1)}{2} \\
&\times&\int_{-1}^{1} d\mu_k \,\left[ \Delta P^{\textrm{obs}}(k,\mu_k) \right]^2 \, \mathcal{P}_l(\mu_k) \, \mathcal{P}_m(\mu_k). \nonumber
\eea

It is possible now to calculate the diagonal covariance matrix for the difference operator $\mathcal{D}$ of Equation\,(\ref{eq:defDk}):
\be
\mathrm{Cov}\mathcal{D}(k)=\sum_{lm} \frac{\partial \mathcal{D}(k) }{\partial P^{\textrm{obs}}_l(k)} G_{lm}(k) \frac{\partial \mathcal{D}(k) }{\partial P^{\textrm{obs}}_m(k)}.
\ee
The needed partial derivatives are
\be
\frac{\partial \mathcal{D}(k)}{\partial P^{\textrm{obs}}_0(k) } =-h_2(\beta,\sigma_v k)
\ee
and
\be
\frac{\partial \mathcal{D}(k)}{\partial P^{\textrm{obs}}_2(k) } =h_0(\beta,\sigma_v k).
\ee

\begin{figure}[!t]
\centering
\includegraphics[scale=0.67]{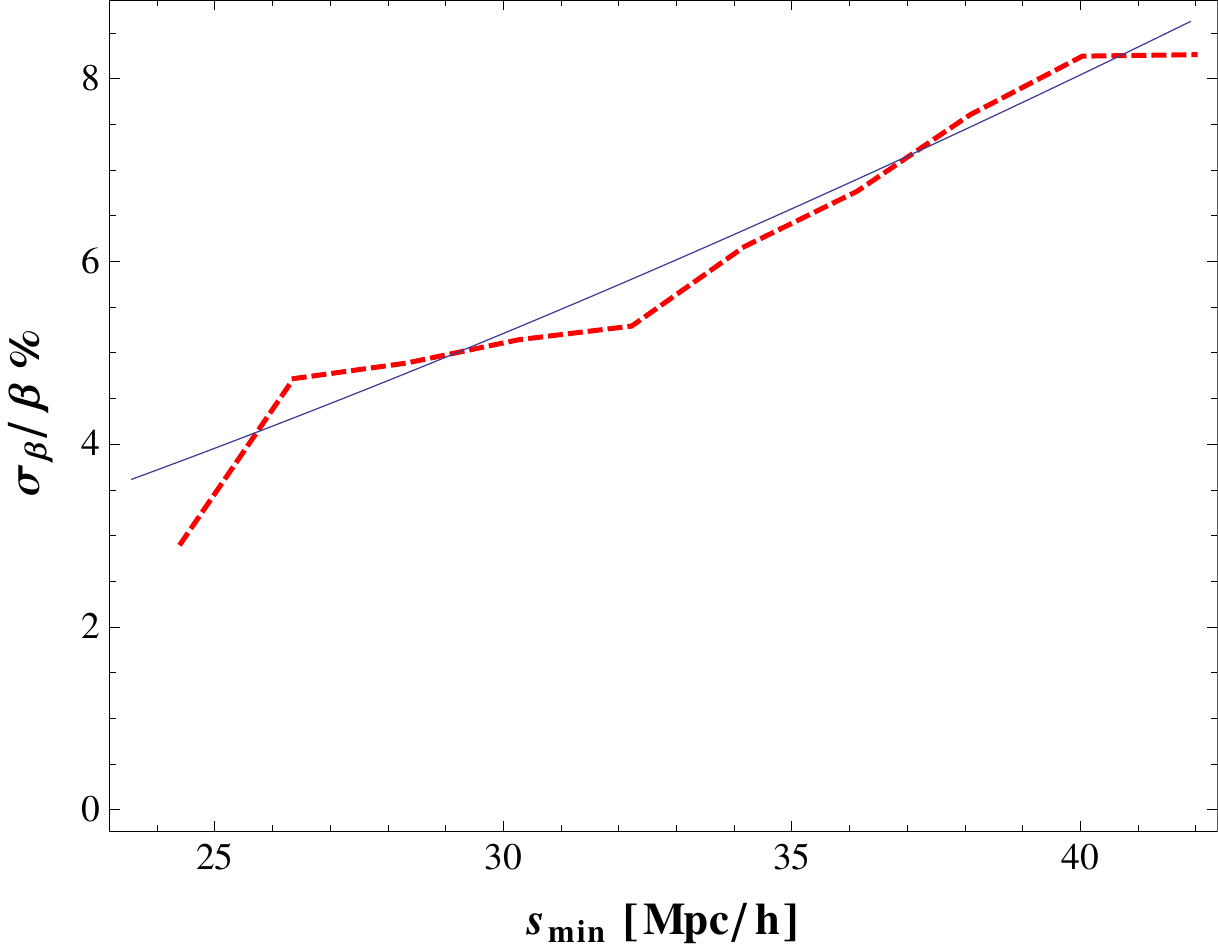}
\caption{\label{fig:fisher}
We show the agreement between the percent fractional marginalized errors on $\beta$ resulting from the $N$-body analysis (dashed thick curve) and a Fisher matrix study (solid thin curve) carried out in Fourier space and having a maximum wavenumber scaled as $k_{\mathrm{max}}=1.5 \pi / s_{\mathrm{min}}$. 
The Fisher matrix method was based on the quantity $\mathcal{D}$ from Equation\,(\ref{eq:defDk}) and refers to the $\Lambda$CDM cosmology used in the simulations and is based on a dark matter survey at redshift $z=1$, of volume $1\,(h^{-1}\,\mathrm{Gpc})^3$, full sky, with negligible shot noise.
}
\end{figure}

Finally, the Fisher matrix for $\beta$ and $\sigma_v$ is specified by
\be
F_{ab}=\sum_{k} \frac{\partial \mathcal{D}(k)}{\partial p_a } \mathrm{Cov}\mathcal{D}(k)^{-1} \frac{\partial \mathcal{D}(k)}{\partial p_b },
\ee
where $p_a=\beta,\sigma_v$ indicates the parameter vector and the partial derivatives can be taken either numerically or in some cases analytically (if for example the distribution $f(v)$ is either a Gaussian or an exponential). We use as limits of the sum $k_{\mathrm
{min}}=2\pi/(V_s)^{(1/3)}$ and utilize a variable maximum wavenumber $k_{max}$ to test the magnitude and trend of the errors.

In Figure\,\ref{fig:fisher} we show the percent fractional error for marginalized errors on $\beta$ for a survey centered at redshift $z=1$, of volume $V_s=1\,\left( h^{-1}\,\mathrm{Gpc} \right)^3$, full sky, with enough dark matter particles to dominate shot noise.
Our Fisher matrix study is based on Fourier space and to relate it to configuration space we need to link wavenumbers to distances. We find that the scaling $k_{\mathrm{max}}=2 \pi/r_{\mathrm{min}}$ gives the Fisher errors of similar magnitude and trend to the errors estimated from the simulations. While a scaling $k_{\mathrm{max}}=1.5 \pi/r_{\mathrm{min}}$ assures that the magnitude and the trends of both the errors are closely matched. We note that a similar scaling $k_{\mathrm{max}}\approx \pi/r_{\mathrm{min}}$ was chosen by tuning in \cite{2011ApJ...726....5O} and \cite{2011MNRAS.417.1913R} to provide a precise link between Fourier and configuration space.

The constraint on $\beta$ can be used to extract information on the growth function if the galaxy linear bias is measured independently. For instance, the amplitude of the matter fluctuations quantified by $\sigma_8^{m}(z=0)$ and measured from a cosmic microwave background radiation experiment could be first scaled to $z=1$, and then the bias could be estimated by $b=\sigma_8^{g}(z=1)/\sigma_8^{m}(z=1)$; the superscripts $m$ and $g$ indicating, as usual, matter and galaxies.  This procedure was followed recently, for example, in \cite{2008Natur.451..541G}, where the fluctuation amplitude derived by the \emph{Wilkinson Microwave Anisotropy Probe experiment} \citep{2007ApJS..170..377S} was utilized.

\section {Conclusion} \label {sec:concl}

We based our analysis on basic RSD streaming models, frequently used as benchmark models in the literature, and often shown, for some galaxy groups, to work reasonably well on quasi-linear scales.
 
We have introduced a new method to determine the RSD $\beta$ parameter in configuration space on quasi-linear scales. The statistics $D$ of Equation\,(\ref{eq:estD}), presents the following advantages: it is not necessary to specify the linear theory predictions for the correlation functions; there is no dependence on the mean galaxy density; only convolutions are involved, assuring stability; it does not rely on galaxy linear bias, apart from the implicit dependence within $\beta$. Random errors on the determinations of the galaxies redshifts are effectively incorporated in the model. 

The method can actually be considered more general than the streaming model because it is insensitive to a factor multiplying the power spectrum that depends only on the wavenumber, which could arise from a quasi-linear correction and/or a scale-dependent galaxy bias. Furthermore perturbation theory might have better convergence properties in configuration space, where our estimator operates, compared to Fourier space \citep{2008MNRAS.390.1470S,2008PhRvD..77f3530M,2009PhRvD..80l3503T}.

In addition, we have noticed a baryonic feature at about $100\,h^{-1}\,\mathrm{Mpc}$ in the normalized quadrupole in configuration space, that should not have a major impact on the extraction of $\beta$ with our method.

We have also carried out analyses based on the $N$-body simulations of \cite{2011PhRvD..84d3501S} and the Fisher matrix method, finding that errors of a few percent on $\beta$ are feasible with a full sky, $1\,(h^{-1}\,\mathrm{Gpc})^3$ survey centered at a redshift of unity and with negligible shot noise and that for minimum separations such that $s_{\mathrm{min}}>35\,h^{-1}\mathrm{Mpc}$ there is no evidence for a bias in the estimation of $\beta$.

It is going to be crucial to be able to measure and interpret in detail the growth of fluctuations, because the information contained therein could indicate deviations from general relativity and constrain the cosmic expansion history. Since subtleties in the systematics of cosmological observations do not guarantee that the theoretical equivalence of Fourier and configuration space can be easily reproduced in practical observations, it should be convenient to have the widest possible arsenal of methods, involving both points of view.

\begin{acknowledgments}
We are thankful to Joe Silk for reading and commenting the manuscript. We are particularly grateful to Masanori Sato for having provided us with the $N$-body simulations results used in this work and for discussions on their features.
\end{acknowledgments}

\bibliographystyle{apj}

\end {document}